# Astro2020 Science White Paper
# Big Science with a nUV-MidIR Rapid-Response 1.3m telescope at L2

**Thematic Areas:** ☐ Planetary Systems  ☒ Star and Planet Formation
☒ Formation and Evolution of Compact Objects  ☒ Cosmology and Fundamental Physics
☒ Stars and Stellar Evolution  ☒ Resolved Stellar Populations and their Environments
☒ Galaxy Evolution  ☒ Multi-Messenger Astronomy and Astrophysics

**Principal Author:**
Name: Jonathan Grindlay
Institution: Center for Astrophysics | Harvard and Smithsonian
Email: jgrindlay@cfa.harvard.edu
Phone: 617-495-7204

**Co-authors:** Edo Berger[1], Brian Metzger[2], Suvi Gezari[3], Zeljko Ivezic[4], Jacob Jencson[5], Mansi Kasliwal[5], Alexander Kutyrev[6], Chelsea Macleod[1], Gary Melnick[1], Bill Purcell[7], George Rieke[8], Yue Shen[9], Nial Tanvir[10], Michael Wood Vasey[11]

[1]Center for Astrophysics | Harvard and Smithsonian, USA, [2]Columbia University, USA, [3]University of Maryland, USA, [4]University of Washington, USA, [5]Caltech, USA, [6]NASA/GSFC [7]Ball Aerospace, USA, [8]University of Arizona, USA, [9]University of Illinois, USA, [10]Leicester University, UK, [11]University of Pittsburgh, USA

**Abstract**: Time-domain Astrophysics (TDA) , a foundation of Astronomy, has become a major part of current and projected (2020's) astrophysics. While much has been derived from temporal measures of flux and color, the real physics comes from spectroscopy. With *LSST* coming on line in 2022, with TDA one of its original drivers, the deluge of Transients and new types of variables will be truly astronomical. With multi-wavelength targeted EM surveys and multi-messenger (e.g. LIGO-international and super ICECUBE), and the possibility of full-sky/full-time X-ray imagers, the discovery of new Transients and Variables will flood telescopes on the ground and in space, and this just for multi-band imaging without spectroscopy. In this White Paper we briefly summarize several long-standing major science objectives that can be realized with TDA imaging and spectroscopy (nUV – mid-IR) from space. We provide a brief description of how these can be achieved with the Time-domain Spectroscopic Observatory (*TSO*), a Probe-Class mission concept that ELTs on ground and Flagship missions in space can not achieve on their own.

**A partial list of Big Science questions still unanswered but relevant to Astro2020**
While we can question what are the truly Big questions, surely we can list a few current candidates and point to the Astro2020 White Papers (WPs) where they are discussed:

1. *Can we directly detect the very first (Pop III) stars with JWST?* The answer is maybe, but only as a population of their resulting (from core collapse) stellar mass black holes (BHs) accreting from binary companions and even then only by observing the strong lensing magnification of such a population in a primordial galaxy by a foreground massive cluster that might require a decade or more of monitoring (Windhorst et al. 2018). Even in the unlikely case that collapse of a massive Pop III star produced a supernova (e.g. from core bounce from an intermediate neutron star), its AB magnitude would be far fainter than *JWST* could detect. As outlined in the Tanvir et al 2019 WP *GRBs as Probes of the Early Universe with TSO*, direct detection of a single (or many) Gamma-ray Burst (GRB) from the inevitable core collapse of PopIII stars is feasible given the rapid response and nearly full-sky response (to full-sky GRB triggers; see forthcoming Grindlay et al *4piXIO* WP) of a 1.3m telescope at L2 with imaging and spectroscopy over the 0.3 - 5µm band. The *Time-domain Spectroscopic Observatory (TSO)* is a Probe-class Mission Concept that is briefly summarized below. *TSO* may be the *only way to detect the existence of what surely would be a Pop III star if observed as a GRB afterglow at redshift z ≥15, a comfortable lower limit on the first Pop III stars.*

2. *Can we use GRBs at z > 6 – 10 to map the Epoch of Reionization (EOR)?* This could be done by using their luminous afterglow emission to map the ionization of the host galaxy vs. local IGM as outlined by McQuinn et al. 2008. The answer is yes, but only if we have at least a modest aperture (e.g. 1.3m) cold telescope in space with imaging and spectroscopy (for prompt identification and followup) that extend to the mid-IR. Such a telescope must be able to *rapidly slew* to a large fraction of the sky, ideally ~90% which is possible from L2 as proposed for *TSO*. Within ~0.5 days of the GRB, the afterglow is bright enough (AB <23) that the GRB redshift and deep diagnostic spectroscopy (for the EOR) can be done. *JWST* cannot do this for every optically dark GRB, and certainly not within ~0.5 days. The past 14 years of GRB observations from the pioneering Neil Gehrels *Swift* Observatory have detected 9 GRBs at z >6 from a sample of ~28% of ~1200 GRBs that are optically dark. Many of these are self-absorbed by dust in their star-forming regions, but those with low NH in their X-ray afterglow spectra are at z >7. These have not been possible to follow up (weather, scheduling) with 8 - 10m telescopes. Typical magnitude limits for J, H, K photometry are <21.5, even from 8 -10m telescopes, due to the bright OH backgrounds. The Tanvir et al WP shows this too can be done with the large sample, possibly ~100/year (full-sky) for GRBs at z >6. *GRB afterglow spectra with TSO would enable multiple sight-lines for exploration of the clumpiness of the EOR as well as its evolution with z back to Pop III* (Question 1, above).

3. *Can we finally understand the origin of the r-process elements?* The recent discovery of the LIGO event GW170817 and its simultaneity with a short Gamma Ray Burst showed conclusively that these are neutron star binary mergers. The prompt and followup emission also showed that the "Kilonova" event that followed was rich with lanthanides indicative that the enormous neutronization from the merger produced significant r-process element production, thereby answering a decades long endeavor to explain their abundance. The Metzger et al 2019 WP shows that these events could be detected out to ~10X distance with the imaging and spectroscopic sensitivity of *TSO*, thereby opening up the exploration space for r-process production well beyond the current LIGO sensitivity limits. In fact LIGO may not be required: short GRBs can be located to ~10arcsec by future wide-field hard X-ray imagers such as the proposed *4piXIO* concept so that immediate followup nUV-mid-IR imaging and spectroscopy with *TSO* can immediately discover and study Kilonovae out to ~1Gpc distances for sufficiently long exposure times.

4. *Can we use the "classic" Reverberation Mapping (RM) technique to measure SMBH masses back to the first Quasars at z >7?* As shown in the Shen et al *Mapping the inner structure of quasars with Time-domain Spectroscopy* WP, this can be done with a 1.3m telescope at L2 and the broad-band coverage (0.3 - 5μm) for imaging and spectroscopy (R = 200 and 1800) which would enable Hbeta RM to be done and SMBH masses derived for Quasars back to z = 8. This is done by repeated spectra of a known quasar at z >7, over months and years, following a flare trigger discovered in the dense sky coverage of that quasar if in the ~2/3 sky monitored by *LSST*. The same measurements can be made on any obviously flaring known quasar (at given z) or any object with colors that prompt spectroscopy to confirm it is a quasar. This permits selection of a large sample of quasars over a range of z to be followed (with *LSST*) for RM measures its SMBH mass. This will enable studies of SMBHmass(z) and also the details of the BLR(z) evolution, which in turn probe dependences on metallicity (Z) at z >3. Once again, the same 1.3m telescope at L2 provides the required deep multi-band imaging and both low and high resolution spectroscopy.

5. What is the nature of the mysterious Red Transients discovered in the Spitzer InfraRed Intensive Transients Survey (SPIRITS) as summarized in Kasliwal et al 2017? These included 64 very luminous transients (SPRITES) so red they with no optical counterparts (Jencson et al 2019). These and other challenges posed by the poorly explored IR Transient sky are summarized by the Kasliwal et al 2019 WP.

Many more examples of high priority TDA science can be cited that *TSO* could best address, but let me conclude with a brief summary of what a *TSO* Probe-class mission would include.

**Brief description of the proposed Probe-Class Time-domain Spectroscopic Observatory (*TSO*)**
*TSO* was proposed as a Probe-class mission concept for study in the 2016 proposal call but not selected, in part because its predecessor the InfraRed Telescope on the *EXIST* mission proposed for Astro2010 had been studied. *TSO* is now a streamlined and easier to build (as Probe) than the IRT. The telescope and focal plane parameters are given in Table 1. The telescope is designed with significant baffling to permit maintaining the radiatively-cooled primary, secondary mirrors and optical bench at T = 110K for pointings ≥30° from the Sun. Slews of up to 180° can be done in ≤8 min to permit rapid acquisition of GRBs. The telescope would be

| **Table 1:** Telescope | 1.3m R-C |
|---|---|
| FoV (arcmin) | 8 x 8 |
| Imaging detectors | 2x2 H2RG |
| Pixel size (arcsec) | 0.25 |
| Imaging/Spectra bands | 4 (parallel) |
| Bands(μm): 0.3-0.73-1.38-2.63-5.0 ||

placed at L2 for 90% sky access (or in a Geosync orbit for 80%), with provision for rapid command uploads for either.

The schematic layout of the focal plane is shown in Fig. 1. The incoming telescope beam is directed to either low-resolution (R = 200) IFU optics to provide spectra in an array of 10 x 10 pixels (0.25") around the central object, OR, if target is bright enough (AB <23) to a high resolution grating for R = 1800 spectra. Images in each of the 4

| Table 2. *TSO* AB mag sensitivities vs. $T_{exp}$ | | | |
|---|---|---|---|
| Operating Mode vs. $T_{exp =}$ | $10^3$ s | $10^4$ s | $10^5$ s |
| Imaging (*each of 4 bands*) | 25.5 | 27 | 28 |
| IFU (R = 200) spectra ( " ) | 22 | 24 | 25.5 |
| Slit (R = 1800) spectra ( " ) | 19.8 | 21.7 | 23 |

bands (Table 1) between 0.30 – 5.0μm are obtained in the 4 H2RG image planes as shown for one such band in Fig. 2. This provides a very powerful combination of imaging and spectroscopy that extends from the nUV to mid-IR, as needed to carry out the science objectives of *TSO*. AB mag sensitivities vs. exposure time for imaging vs. spectroscopy are given in Table 2.

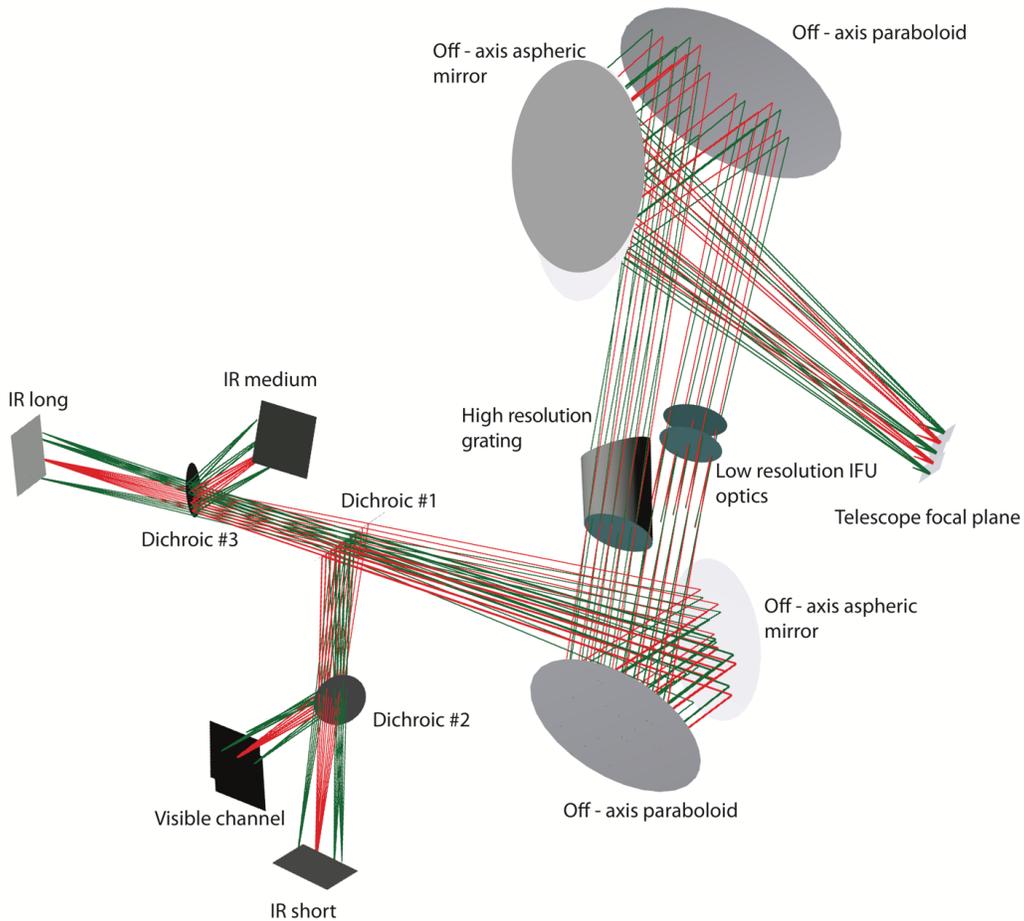

**Fig. 1.** Optical design of the *TSO* focal plane for imaging and either IFU (R = 200) or grating (R =1800) spectroscopy in each of 4 bands (simultaneous) in Table 1.

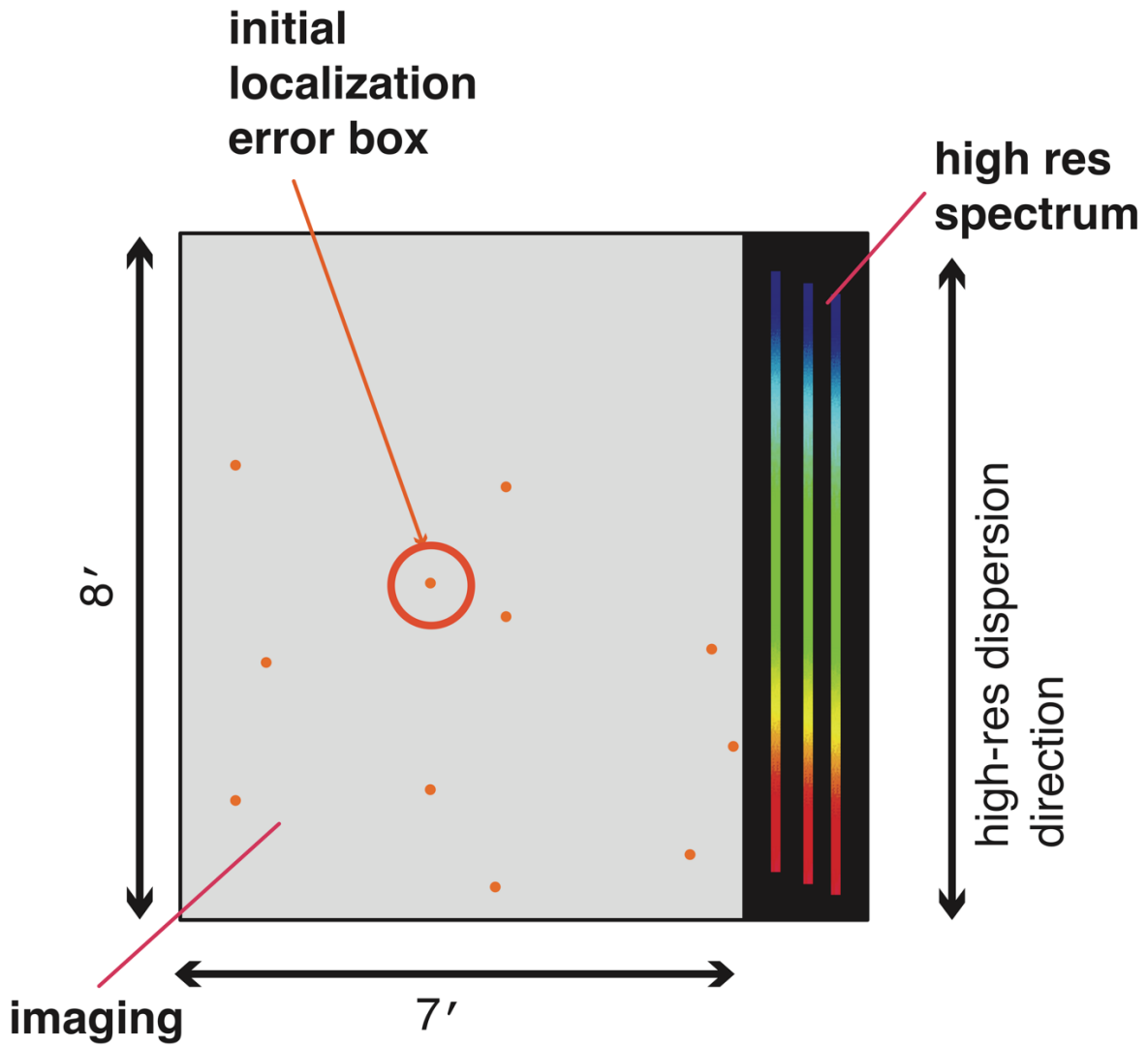

**Fig. 2.** Imaging plane vs. high resolution spectroscopy allocation of the H2RG detector plane for *TSO*. The IFU readout is on a separate detector for initial low resolution but spatially resolved (10 x 10 0.25" pixels) spectroscopy to yield a <0.25" position (if not already obtained from initial multi-band images) for the GRB or target to enable hi-res spectroscopy with a 0.5" slit.

Full details for the *TSO* Science, Mission and Operations will be provided in the Mission WP submission.